\documentclass[twocolumn,showpacs,preprintnumbers,amsmath,amssymb]{revtex4}
\usepackage{graphicx}
\usepackage{dcolumn}
\usepackage{bm}
\def\al13co4{{Al$_{13}$Co$_{4}$}}

\begin{document}

\preprint{}

\title{Structure and stability of Al$_2$Fe}

\author{M. Mihalkovi\v{c}}
\affiliation{Institute of Physics, Slovak Academy of Sciences, 84228 Bratislava, Slovakia}
\author{M. Widom}
\affiliation{
Department of Physics, Carnegie Mellon University\\
Pittsburgh, PA  15213
}

\date{\today}
\begin{abstract}
We employ first principles total energy and phonon calculations to address the structure and stability of Al$_2$Fe.  This structure, which is reported as stable in the assessed Al-Fe phase diagram, is distinguished by an unusually low triclinic symmetry.  The initial crystallographic structure determination additionally featured an unusual hole large enough to accommodate an additional atom.  Our calculations indicate the hole must be filled, but predict the triclinic structure is unstable relative to a simpler structure based on the prototype MoSi$_2$.  This MoSi$_2$ structure is interesting because it is predicted to be nonmagnetic, electrically insulating and high density, while the triclinic structure is magnetic, metallic and low density.  We reconcile this seeming contradiction by demonstrating a high vibrational entropy that explains why the triclinic structure is stable at high temperatures.  Finally, we note that Al$_5$Fe$_2$ poses a similar problem of unexplained stability.
\end{abstract}

\maketitle

\section{Introduction}
Aluminum-based intermetallic alloys with transition metals are of high interest for their complex crystalline and quasicrystalline structures, formed primarily with late transition metals, and their technologically useful compounds formed primarily with early transition metals.  Experimental phase diagram determination is difficult because many phases often exist within small composition ranges, many structures have unusually large unit cells and many are intrinsically disordered, exhibiting mixed or partial site occupancy.  First principles calculations can help resolve some uncertainties in the phase diagrams, but are challenging themselves, for many of the same reasons.  Intrinsic disorder requires studying alternative realizations of specific site occupancy.  Some of the nearby competing phases may have unknown or poorly known structures.  The large unit cells pose computational difficulties.  Further complicating the study is the prevalence of magnetism among late transition metals.

Although specific Al-Fe phases are not of direct commercial or technological interest (FIX THIS CLAIM!!!), precipitates of Al-Fe compounds can enhance the high temperature strength of pure Al.  Al-Fe is also the prototype binary magnetic magnetic alloy based on a bcc structure~\cite{Solorio88}.  Our own interest in the Al-Fe phase diagram derives from its complex and disordered crystal structures, some of which are related to quasicrystals.

The compound Al$_2$Fe is of special interest because of its unusual lowest-possible symmetry crystal structure, triclinic with space group \# 1 (P1) as determined by Corby and Black~\cite{Corby73}.  The initial crystallographic refinement, based on anomalous dispersion experiments, proposed an 18-atom unit cell (Pearson symbol aP18) with an unusual ``hole'' that was sufficiently large to fit an entire Al or Fe atom.  They also reported three sites of mixed occupancy, Al$_{0.5}$Fe$_{0.5}$.

Our preliminary first principles calculations of total energy~\cite{alloy05} showed that filling the hole was energetically favorable, thus we predicted the correct Pearson type as aP19~\cite{alloy05}. However, we found this structure to be unstable with respect to competing phases of differing composition, no matter how the hole was filled and how the partial occupancy was resolved.  An alternate structure based on the MoSi$_2$ prototype (Pearson symbol tI6) was predicted to be the true stable structure.  This tI6 structure can be considered as an Al-rich variant of the B2 (Pearson cP2) structure of AlFe.  It has never been observed experimentally, although it would be of high interest because it is predicted to be electrically insulating with a narrow gap~\cite{Weinert98,Krajci02}.  Instead, multiple reexaminations confirm the stability of a triclinic structure for Al$_2$Fe.  A recent crystallographic refinement~\cite{Chumak10}, utilizing conventional single crystal diffraction, proposes the space group is \# 2 (P$\bar{1}$) and fills the hole, confirming our predicted Pearson type aP19.

Here we present a thorough study of the stability of Al$_2$Fe utilizing first principles total energy calculations of low temperature enthalpy supplemented by a phonon-based calculation of vibrational entropy yielding the high temperature Gibbs free energy.  Our calculations predict that the tI6 structure loses stability to the aP19 structure at elevated temperatures.  This occurs because aP19 has a much lower atomic density than tI6, resulting in high vibrational entropy.  A similar stabilization effect due to vibrational entropy was observed in the $\theta/\theta'$ system of Al$_2$Cu~\cite{Wolverton01}.

The remainder of this introduction surveys the global Al-Fe phase diagram and presents our calculational methods.  We then present a thorough investigation of the energetics of plausible aP18 and aP19 structures, including the effects of magnetic moment formation and antiferromagnetism.  Finally we present vibrational densities of states that display a large enhancement of low frequency phonons in the aP19 structure relative to tI6.

\subsection{Assessed Al-Fe Phase Diagram}
The Al-Fe phase diagram~\cite{Sundman09} contains at least six compounds as well as the two pure elements.  Additionally there are at least three known metastable phases.  Table~\ref{tab:phases} displays pertinent information including names, composition ranges, Pearson types, space groups and assessed stability of all reported phases.  Several of the phases report composition ranges associated with chemical substitution between Al and Fe and also partial site occupancy.  The Al$_3$Fe phase, more accurately described as Al$_{13}$Fe$_4$, is well known as a decagonal quasicrystal approximant.  Structures of Al$_2$Fe and Al$_5$Fe$_2$ also feature pentagonal networks~\cite{Hirata08}.

\begin{table}[h]
\begin{tabular}{lr|lr|lrr}
Name         &  \% Fe& Pearson & Group      & Stab & $\Delta H_{\rm for}$ & $\Delta E$ \\
\hline
Al           &     0 & cF4          & Fm$\bar{3}$m       & S  &    0 &   NA \\
Al$_6$Fe     &    14 & oC28         & Cmc2$_1$           & M  & -205 &  1.4 \\
Al$_9$Fe$_2$ &    18 & mP22         & D8$_d$             & M  & -258 &  9.5 \\
Al$_3$Fe     & 23-26 & mC102        & C2/m               & S  & -347 &   NA \\
Al$_5$Fe$_2$ & 27-30 & oC24         & Cmcm               & S? & -349 &  5.7 \\
Al$_2$Fe     & 33-34 & aP18(19)     & P1($\bar{1}$)      & S? & -337 & 29.1 \\
$\epsilon$-Al$_8$Fe$_5$& 35-42& cI52& I$\bar{4}$3m       & HT & -286 & 74.6 \\
AlFe         & 45-77 & cP2          & Pm$\bar{3}$m       & S  & -346 &  NA  \\
AlFe$_2$     &    67 & cF24         & Fd$\bar{3}$m       & M  & -116 &131.1 \\
AlFe$_3$     & 66-77 & cF16         & Fm$\bar{3}$m       & S  & -198 &   NA \\
Fe           & 55-100& cI2          & Im$\bar{3}$m       & S  &    0 &   NA \\
Fe           & 98-100& cF4          & Fm$\bar{3}$m       & HT &  +80 & 80.2 \\
\end{tabular}
\caption{\label{tab:phases} Known phases of Al-Fe.  Stability designation: S=stable to low T; S?=stable at high temperature down to unknown T$<$400C; M=metastable; HT=stable at high temperature only.  See text section~\ref{sec:methods} for $\Delta H_{\rm for}$ and $\Delta E$ (units are meV/atom).}
\end{table}

\subsection{Methods}
\label{sec:methods}
Our calculations follow methods outlined in a prior paper~\cite{Mihal04}.  We utilize VASP~\cite{Kresse93,Kresse96} to carry out first principles total energy calculations in the PW91 generalized gradient approximation.  Comparisons with the LDA and PBE density functionals confirm the principal findings based on PW91.  We relax all atomic positions and lattice parameters, and increase our k-point densities until energies have converged.  We adopt projector augmented wave potentials~\cite{Blochl94,Kresse99} and maintain a fixed energy cutoff of 267.9 eV (the default for Fe).  All calculations considered the possibility of spin polarization, and utilize the ``Medium'' precision setting which allows small wrap-around errors in Fourier Transforms.

Given total energies for a variety of structures, we calculate the enthalpy of formation $\Delta H_{\rm for}$ which is the enthalpy of the structure relative to a tie-line connecting the ground state configurations of the pure elements.  The convex hull of these energies constitute the predicted low temperature stable structures.  For structures that lie above the convex hull we calculate the instability energy $\Delta E$ as the enthalpy relative to the convex hull.  When presented with mutiple structure possibilities, or mixed site occupancy, we examine all plausible structures and report the most energetically favorable.

\section{Results}
\subsection{Global phase diagram}
\begin{figure*}
\includegraphics[width=4in,angle=-90]{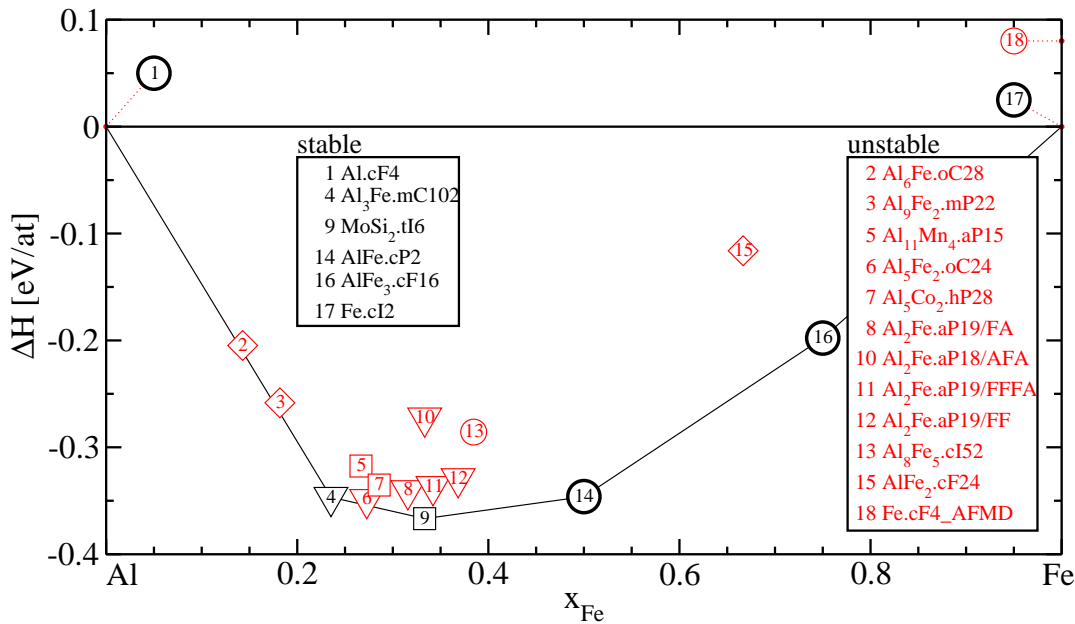}
\caption{\label{fig:global} Enthalpies of formation of Al-Fe compounds.  Plotting symbols: heavy circle=known stable; light circle=known high temperature; diamond=known metastable; triangle=unknown stability; square=unknown phase.  Plotting colors: black=on convex hull; red=above convex hull.  Notation A and F indicate Al or Fe on Wyckoff $1a$ positions M1, M2 and M3 (aP18) or Wyckoff $2i$ position Fe4 (aP19).}
\end{figure*}

As illustrated in Fig.~\ref{fig:global} our calculated enthalpies agree in almost every respect with the assessed phase diagram.  All the phases that are known to be stable at low temperature (heavy circles) indeed reach the convex hull.  Those known to be metastable (diamonds) or stable only at high temperatures (light circles) lie slightly above the convex hull.  Of those whose low temperature stability is uncertain (triangles), Al$_3$Fe lies on the convex hull, while Al$_5$Fe$_2$ and Al$_2$Fe (aP18 and aP19) both lie slightly above.  The only serious descrepancy between the experimental phase diagram and our calculation is the presence of Al$_2$Fe in the tI6 (MoSi$_2$) structure on the convex hull.

Magnetism was found to be favorable in all the structures reported in Table~\ref{tab:phases} containing 33\% or more Fe.  Elemental Fe exhibits ferromagnetism in its low temperature cI2 structure and modulated antiferromagnetism in its high temperature cF4 structure.  Al$_2$Fe also exhibits long-wavelength antiferromagnetism in its aP18 and aP19 structures.

Our calculated total energies correctly predict the Al$_9$Fe$_2$ and Al$_6$Fe structures to be metastable in the Al-Fe alloy system, while the same structures are correctly predicted to be stable in the Al-Co and Al-Mn alloy systems, respectively~\cite{alloy05,Mihal07}. Similarly, we correctly predict Al-Fe to be unstable when placed in the Al$_{11}$Mn$_4$.aP15 and Al$_5$Co$_2$.hP28 structures.  Conversely, we correctly predict AlFe$_3$ stable in Al-Fe but unstable in both Al-Co and Al-Mn.  This sensitivity to the small differences in interatomic bonding between Fe and its neighbors in the periodic table, Co and Mn, gives us confidence in the validity of our first-principles total energies.

\subsection{Al$_2$Fe.aP18/aP19}
The Corby and Black~\cite{Corby73} structure with Pearson type aP18 possesses an unusually large hole.  In addition three sites, labeled M1, M2 and M3, exhibit mixed Al$_{0.5}$Fe$_{0.5}$ occupancy.  We tested all eight arrangements of Al and Fe among these three sites (labeled AAA, AAF, \dots, FFF) within a single unit cell and found that none of them resulted in a stable structure.  Additionally, all suffered rather large maximum atomic displacements of 0.3-0.4~\AA.  All structures except AAA favored weak ferromagnetism. Our optimal stoichiometric structure, AFA, exhibited magnetic moments averaging 1 $\mu_B$/Fe atom, resulting in an energy drop of 26 meV/atom. 

Previously~\cite{alloy05}, we found that filling the Corby-Black hole with an Fe atom is energetically favorable, suggesting that aP19 is the correct Pearson type for this compound.  The nominal composition Al$_2$Fe cannot be achieved in a single unit cell.  Additionally, in the new structure refinenement~\cite{Chumak10}  there is a single site (Fe4, Wyckoff position 2i) with mixed Al$_{0.295}$Fe$_{0.705}$ occupancy.  For this reason we examine both Al and Fe on the Fe4 site within a single unit cell (labeled AA, AF=FA, FF), and we also examine a supercell doubled along the $a$-axis with one of the four Fe4 sites replaced by Al (labeled FFFA).

In our single cell aP19/FF ferromagnetic structure we find moments of: 1.1$\mu_B$ on Fe1; 0.9 on Fe2 sites; 0.8 on Fe3; 1.8 on Fe4.  In the supercell we find long-period antiferromagnetism is favored over ferromagnetism by 3 meV/atom, consistent with the findings of incommensurate antiferromagnetism~\cite{Kaptas06}.  Magnetic moments in our aP19/FFFA antiferromagnetic structure were qualitatively similar in magnitude to our aP19/FF ferromagnetic structure except in the vicinity of Al substitutions on the Fe4 site.  Attempts to find disordered magnetic ground states by starting from randomly selected initial Fe moments of $\pm2\mu_B$ did not succeed in lowering the total energy, suggesting a possible absence of spin glass order~\cite{Ross01}.  Likewise, runs utilizing noncollinear magnetism failed to reduce the energy.

We note a slope in the enthalpies of aP19 with respect to $x_{\rm Fe}$ so that $\Delta E$ drops to as low as 9 meV/atom in the Al-rich limit of aP19/AA (not shown) at $x_{\rm Fe}=0.2632$, while it rises to 33.3 meV/atom in the Fe-rich limit of aP19/FF (reference number 12 in Fig.~\ref{fig:global}) at $x_{\rm Fe}$=0.3684. The optimal structure we found at the precise $x_{\rm Fe}=0.3333$ stoichiometry of Al$_2$Fe is a six-fold supercell of aP19 (not shown) containing 114 atoms and remaining unstable by 29 meV/atom.  The experimentally reported composition is $x_{\rm Fe}$=0.3374.

\section{Vibrational and configurational free energies}

\begin{figure*}
\includegraphics[width=2.5in,angle=-90]{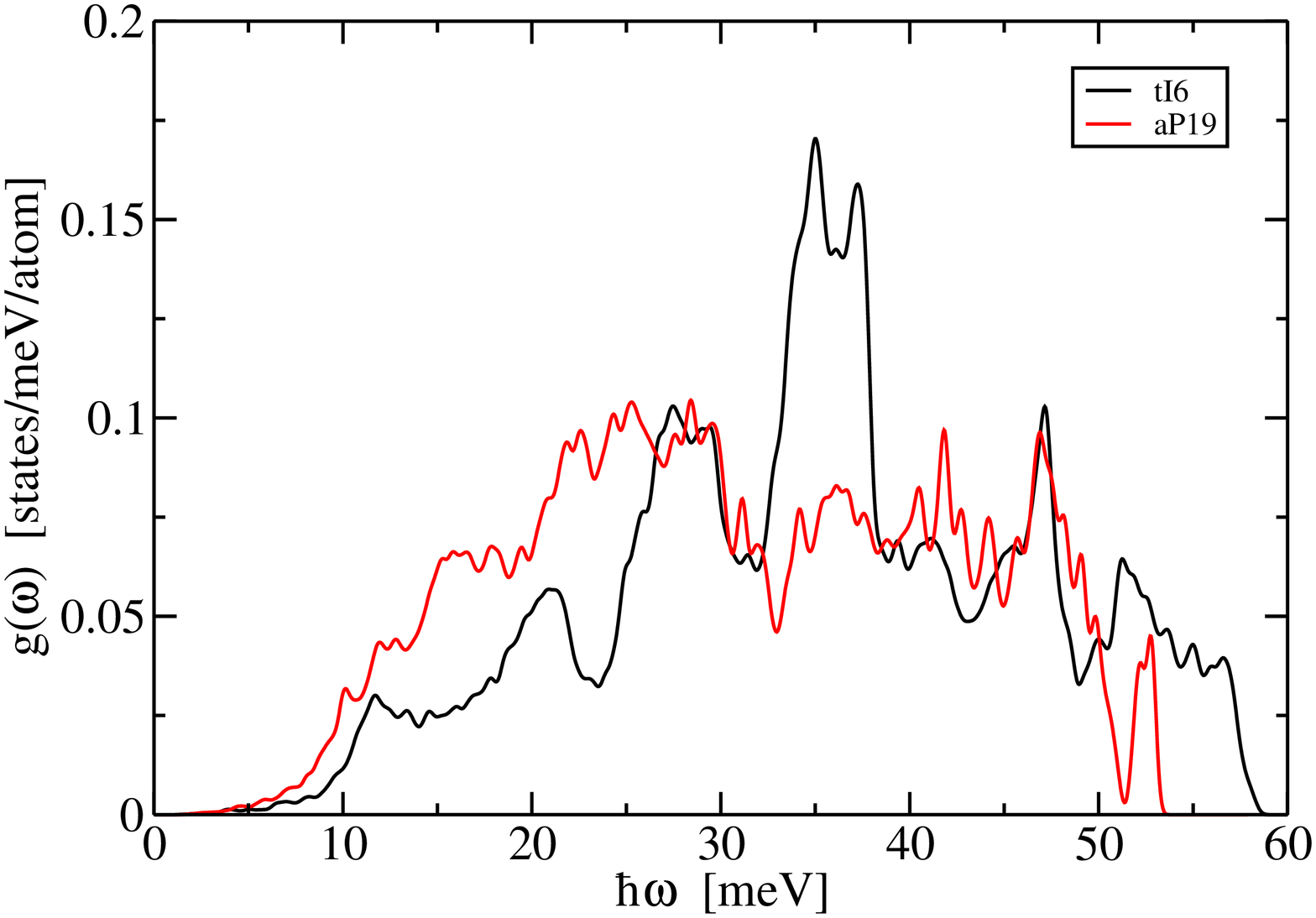}
\includegraphics[width=2.5in,angle=-90]{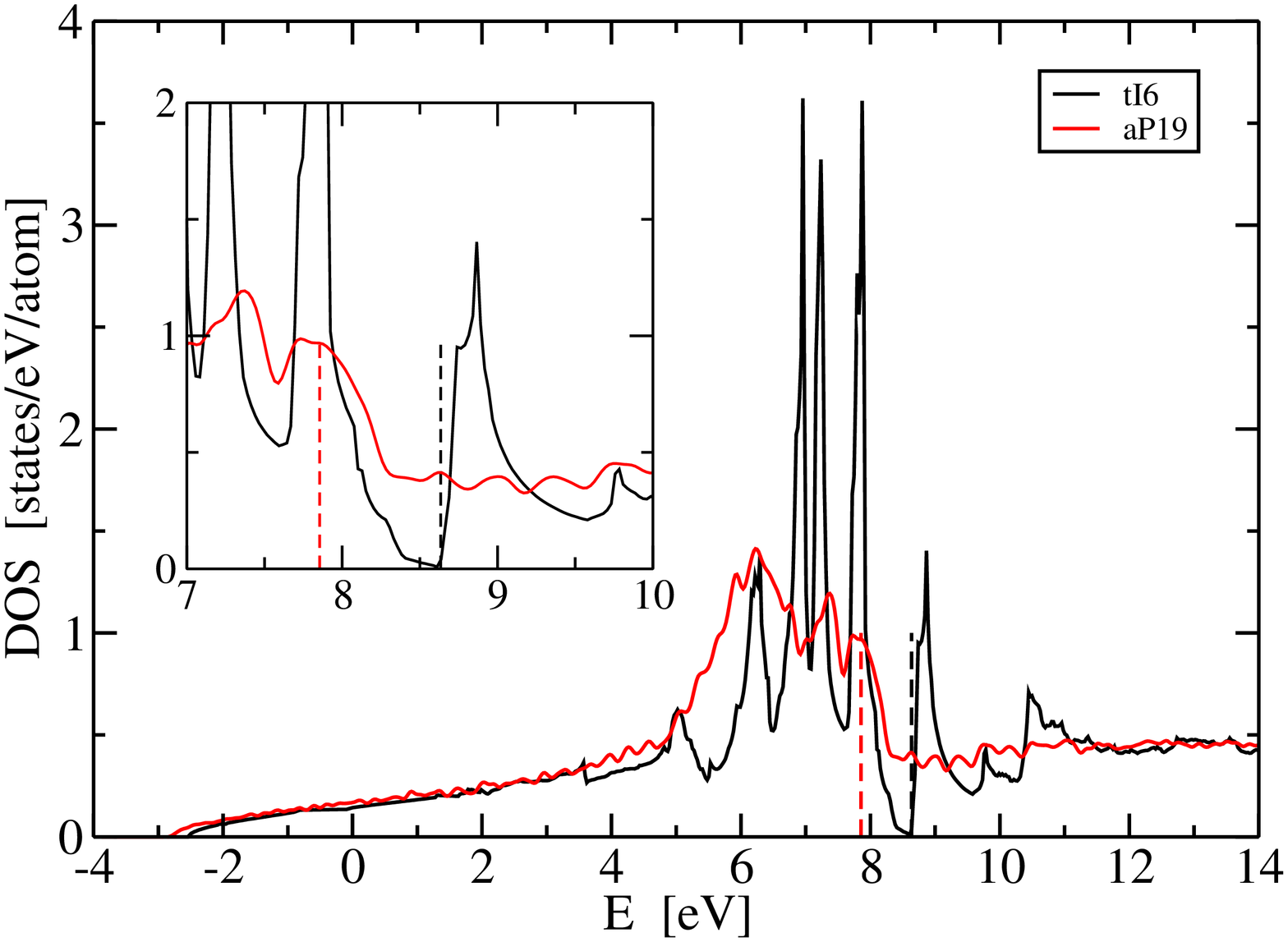}
\caption{\label{fig:dos} (color online) tI6 and aP19/FFFA vibrational (left) and electronic (right) densities of states.  For convenience the electronic DOS of aP19 is shown {\em without} spin polarization.  Vertical dashed lines indicate Fermi energy.}
\end{figure*}

Despite extensive effort, we were unable to find an enthalpy-minimizing triclinic structure for Al$_2$Fe.  All structures considered remained above the convex hull as illustrated in Fig~\ref{fig:global}.  Given that the assessed phase diagram does not assert low temperature stability, we now consider the possibility that this structures is stabilized at high temperature by entropy.  An entropy $s$ per atom results in a free energy reduction of $Ts$ at temperature $T$.  The two primary sources of entropy are continuous vibrational and discrete configurational degrees of freedom.

Even with the hole filled, the atomic volume of aP19/FFFA is 13.48~\AA$^3$/atom while the atomic volume of tI6 is only 12.7~\AA$^3$/atom.  The higher atomic volume of aP19 implies weaker bonding, consistent with lower enthalpy of formation.  Weaker bonding suggests an enhanced density of low frequency phonons that can increase vibrational entropy and correspondingly lower the vibrational free energy.  Because Al$_2$Fe.aP19/FFFA is at very nearly the identical composition to Al$_2$Fe.tI6, we need only compare the vibrational free energies of these two structures (in principle we should include a small admixture of AlFe.cI2, but with {\em very} low weighting).

The vibrational free energy can be obtained within the harmonic approximation, given a vibrational density of states $g(\omega)$, by means of the quantum mechanical theory of phonons as expressed by
\begin{equation}
f_{vib}(T) = k_B T \int g(\omega) \ln{[2\sinh{(\hbar \omega/2 k_B T)}]} d\omega
\end{equation}
To calculate $g(\omega)$ we employed the force constant method for phonon calculations~\cite{Kresse95}, in which we take a supercell of minimum edge length 8~\AA~ and evaluate the second derivatives of total energy $\partial^2 U/\partial {\bf R}_i\partial {\bf R}_j$ for all pairs of atoms $i$ and $j$.  The calculations utilized density functional perturbation theory as implemented in VASP 5.2.  We then construct the dynamical matrix~\cite{Ashcroft} ${\bf D(k)}$ on a dense mesh of $k$-points and evaluate the phonon frequencies.  For our phonon calculations we revise our standard VASP calculations to use ``Accurate'' precision so as to avoid wrap-around errors, and we set EDIFF=5$\times 10^{-10}$.

The resulting calculated vibrational densities of states are presented in Fig.~\ref{fig:dos}.  The excess density of low frequency phonons is clearly evident and yields a reduction in free energy of 38 meV/atom by the time we reach T=1000K.

In addition to the continuous vibrational degrees of freedom, some entropy is contained within the discrete configurational fluctuations associated with the partially occupied state. We estimate this assuming independent occupancies of every mixed or partially occupied site.  For the case of Al$_2$Fe, the mixed Al/Fe occupancy on the Fe4 site yields
\begin{equation}
s = - \frac{2}{19} k_B \left( f_{Al} \ln f_{Al}+ f_{Fe} \ln f_{Fe} \right)
\end{equation}
where the factor of 2/19 arises because there are 2 Fe4 sites/cell and the occupancy factors are $f_{Al}$=0.295 and $f_{Fe}=1-f{Al}$=0.705.  Setting T=1000K we find that mixed occupancy can account for up to 5 meV/atom of stability.  This is an upper bound on the actual value, because correlations in site occupancy will slightly reduce the entropy.

Combining the vibrational free energy and the additional discrete configurational entropy results in a significant lowering of the Gibbs free energy of Al$_2$Fe.aP19.  We find this free energy reaches the convex hull around T=500K.  The competing tI6 phase loses stability by T=550K.

\section{Al$_5$Fe$_2$}

\begin{figure*}
\includegraphics[width=2.5in]{xmfig.eps}
\hspace{.5cm}
\includegraphics[width=2.5in]{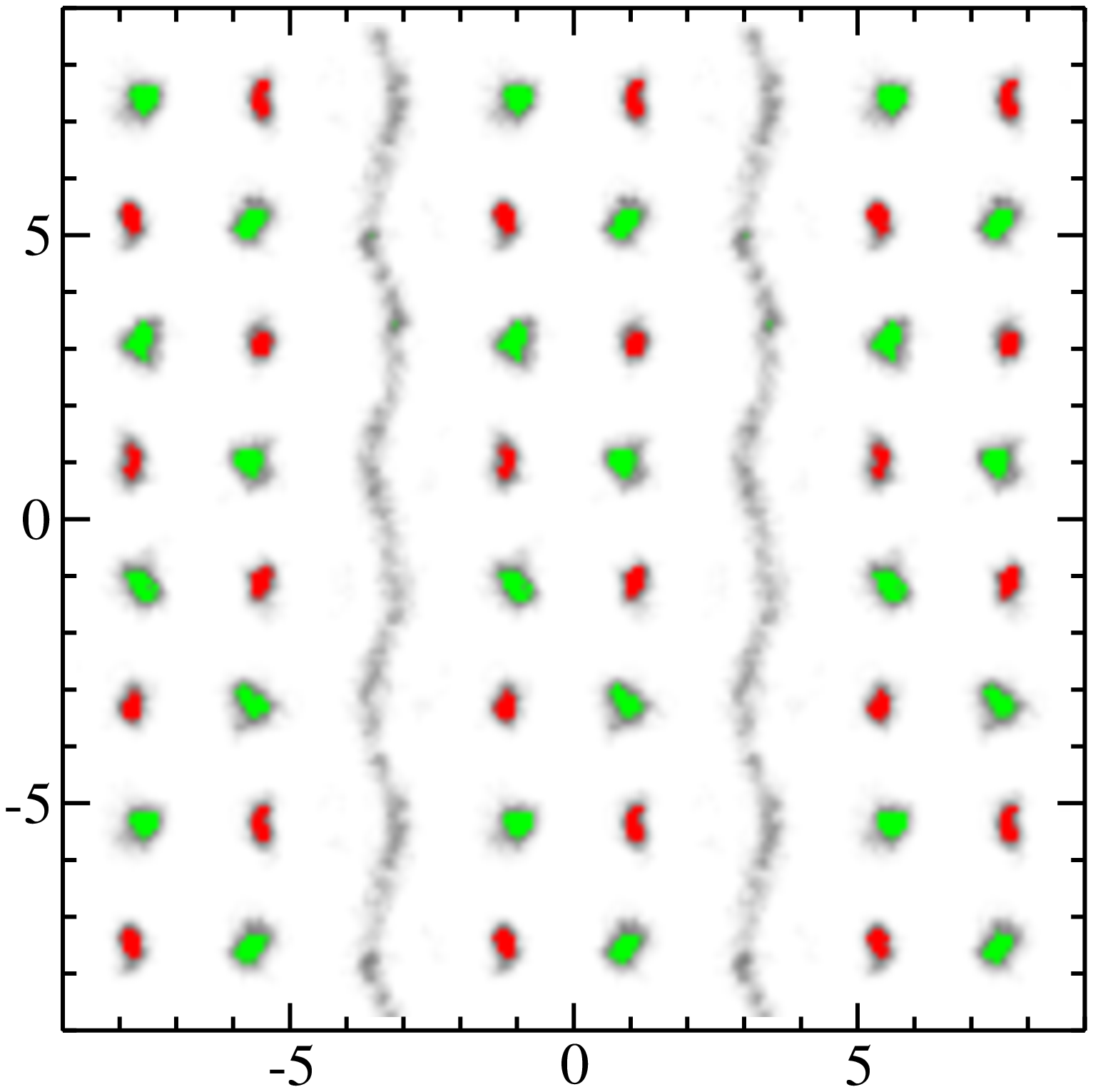}
\vspace{.5cm}
\caption{\label{fig:Al5Fe2} (color online) Structure of Al$_5$Fe$_2$ showing aluminum channels. The $b$-axis is horizontal, and the $c$-axis is vertical. (left) ideal period 3 superstructure. (right) Ab-initio MD at T=1300K.}
\end{figure*}

The other phase of problematic stability is Al$_5$Fe$_2$ (see Fig.~\ref{fig:Al5Fe2}).  As noted previously this structure features pentagonal networks as well as ``aluminum channels''.  Aluminum atoms can diffuse along the channels in a liquid-like fashion, and the density of the channels (i.e. the number of aluminums per $c$-repeat per channel) can vary over a range around one Al/repeat/channel.  We find the energetically optimal structure occurs with four Al atoms in three repeats for a channel density of 4/3.  The additional Al atoms raise the Fermi energy to the vicinity of a strong pseudogap in the electronic density of states, which can account for its energetic optimality.  However, the energy remains high relative to the convex hull by about 6 meV/atom. Preliminary phonon calculations fail to explain the observed high temperature stability of the phase.

We believe that full anharmonic vibrational free energy calculations are needed, owing to the liquid-like motion of the channel Al atoms.  This behavior is evident in Fig.~\ref{fig:Al5Fe2} where we present the occupation densities $\rho_{\alpha}({\bf r})$ using a color scheme where gray represents low density, while red and green represent high density of iron and aluminum, respectively.  The data was collected using VASP molecular dynamics for a duration of 15ps.

\section{Conclusion}

The principal aim of this paper was to explain the experimental observation of Al$_2$Fe in the low symmetry triclinic structure aP19 when a simpler and energetically preferable tetragonal tI6 structure is available.  Since aP19 is stabilized at high temperature by its high vibrational and discrete configurational entropy, it will naturally form first in samples cooled from the melt.  Other high temperature phases at nearby compositions such as Al$_5$Fe$_2$ and Al$_8$Fe$_5$ compete with Al$_2$Fe, likely further frustrating attempts to achieve the tI6 structure through off-stoichiometric growth.  Still, density functional theory predicts the true low temperature structure is tI6, and we encourage experimentalists to search for it.  The payoff would be obtaining an insulating compound consisting of metals Al and Fe.

\acknowledgments
This work was supported in part by grants VEGA j2/0111/11 and APVV-0647-10.

\bibliography{alfe}

\end{document}